\documentclass[manuscript]{acmart}

\AtBeginDocument{%
  }

\setcopyright{acmlicensed}
\copyrightyear{2024}
\acmYear{2024}
\acmDOI{XXXXXXX.XXXXXXX}

\acmConference[CSCW '24]{workshop ``From Stem to Stern: Contestability Along AI Value Chains'' at the Conference for Computer Supported Collaborative Work}{09 November 
  2024}{San José, Costa Rica}
\acmISBN{978-1-4503-XXXX-X/18/06}

\begin{document}

%% Our commands to let us do in-line comments
\newcommand\Mark[1]{\textcolor{blue}{\small [Mark: #1]}}
\newcommand\MarkLeft[1]{\textcolor{blue}{\small $\leftarrow$Mark: #1}}
\newcommand\MarkRight[1]{\textcolor{blue}{\small Mark: #1$\rightarrow$}}

\newcommand\Gennie[1]{\textcolor{red}{\small [Gennie: #1]}}
\newcommand\GennieLeft[1]{\textcolor{red}{\small $\leftarrow$Gennie: #1}}
\newcommand\GennieRight[1]{\textcolor{red}{\small Gennie: #1$\rightarrow$}}

\title{Recognizing Lawyers as AI Creators and Intermediaries in Contestability}

\author{Gennie Mansi}
\email{gmansi3@gatech.edu}
\orcid{0000-0001-6186-5102}
\author{Mark Riedl}
\email{riedl@cc.gatech.edu}
\affiliation{%
  \institution{Georgia Institute of Technology}
  \city{Atlanta}
  \state{GA}
  \country{USA}
}

\renewcommand{\shortauthors}{Mansi \& Riedl}

\begin{abstract}
Laws play a key role in the complex socio-technical system impacting contestability: they create the regulations shaping the way AI systems are designed, evaluated, and used. Despite their role in the AI value chain, lawyers’ impact on contestability has gone largely unrecognized in the design of AI systems. In this paper, we highlight two main roles lawyers play that impact contestability: (1) as AI Creators because the regulations they create shape the design and evaluation of AI systems before they are deployed; and (2) as Intermediaries because they interpret regulations when harm occurs, navigating the gap between stakeholders, instutions, and harmful outcomes. We use these two roles to illuminate new opportunities and challenges for including lawyers in the design of AI systems, contributing a significant first step in practical recommendations to amplify the power to contest systems through cross-disciplinary design.
\end{abstract}

\begin{CCSXML}
<ccs2012>
   <concept>
       <concept_id>10003120.10003121.10003126</concept_id>
       <concept_desc>Human-centered computing~HCI theory, concepts and models</concept_desc>
       <concept_significance>500</concept_significance>
       </concept>
 </ccs2012>
\end{CCSXML}

\ccsdesc[500]{Human-centered computing~HCI theory, concepts and models}
\keywords{algorithmic decision-making, lawyers, cross-disciplinary design, contestability }

\received{15 September 2024}

\maketitle

\section{Introduction}
From design to deployment, the impacts of laws and regulations on the AI life cycle dramatically shape {\em contestability}---how an AI system is open and responsive to human dispute and intervention \cite{KarusalaEtAl2024}.  Prior to deployment, laws and regulations impact how AI systems are created, what algorithms can be selected, and how safety is measured and determined. For example, the Federal Drug Administration (FDA) requires AI-based devices remain ``on leash'' such that developers can always provide a detailed description of how the system will evolve over time, significantly impacting choices around how the kinds of algorithms used in AI systems \cite{FDA_AIMedical}. Regulations also govern required procedures to prevent risks \cite{Chan2021}, train AI systems to meet localized needs \cite{MalihaEtAl2021}, and how their application must be justified and presented to consumers \cite{MulligaEtAl2019}. After deployment, regulations impact
%how AI systems operate as sociotechnical systems because they shape 
how different actors contribute to their deployment and use, further shaping AI systems' sociotechnical environment. Regulations determine procedures for how to complete audits \cite{MulligaEtAl2019, FDA_AIMedical} and for litigating harms once they have been caused \cite{Sundholm2024, PriceAndCohen2023_locating, GerkeEtAl2020}, which can cause users' to change or adopt new behaviors as they try to mitigate perceived legal risks from the system \cite{NashEtAl2004}. Consequently, laws and regulations influence how contestability is enacted across the sociotechnical environment, directly shaping AI systems in ways already recognized as critical to contestability \cite{AlfrinkEtAl2023_Framework}.

Although laws and regulations play a significant role in how contestability is enacted throughout the AI life cycle, there is limited recognition for lawyers as the actors who create these legal structures.
Lawyers, then, are people who, through the regulations they create, directly shape how stakeholders are empowered to prevent and recover from AI system harms. Therefore, they are significant actors who should be included when designing for contestability. \textbf{There is a need to investigate how to include lawyers as a part of design efforts for contestable AI systems.}
However, there is limited understanding around the roles lawyers play in creating the legal context for AI systems. More recent work in the Contestability Community has begun to interview lawyers, bringing attention to the need to design with them in order to improve AI systems' contestability \cite{JinAndSalehi2024, WarrenAndSalehi2022}. Consequently, there is a need to understand the capacities through which we can understand, communicate with, and leverage lawyers' perspectives and skills when creating AI systems.

\textbf{Towards this end, our work contributes a first step in elucidating specific ways lawyers impact the AI value chain and identifying opportunities for cross-disciplinary design.}
In this paper, we argue that recognizing how lawyers shape the contestability of AI systems through two key roles---(1)~as \textit{AI Creators} and as (2)~\textit{Intermediaries}---allows us to understand how to direct efforts to design AI systems {\em with} them. We discuss how designing for lawyers as AI Creators and lawyers as AI Intermediaries can better enable the development of contestable systems. We also present anticipated challenges and opportunities when designing AI systems with lawyers in these two roles. We will predominantly use AI-based medical technology for examples of the legal terms. However, these principles are present in other domains as well, including finance, criminal justice, and privacy among others \cite{AlfrinkEtAl2023_Framework, KarusalaEtAl2024}. Through our work we aim to include and provide practical recommendations to leverage the experiences and knowledge of lawyers as actors who powerfully influence the implementation of contestability.

\section{Lawyers as AI Creators}
Lawyers act as \textit{AI Creators} because the laws and regulations impact AI systems' designs, including mechanisms for contestability. Lawyers create laws to directly shape the design and deployment of AI systems to protect consumers. AI is what lawyers call a ``credence good''---a consumer can't know the quality of the good until after the purchase. It requires blind faith in the quality or some rigorous, systemic evaluation in order to ensure it works properly \cite{Price2022}. 

Due to the risk consumers assume with AI as a credence good, lawyers regulate its development at multiple points through \textit{ex ante} and \textit{ex post} laws. 
\textit{Ex ante} laws  impact technology design before it enters stakeholders' workflows. For example, current FDA regulations ensure and validate the safety of technology before physicians' use. 
\textit{Ex post} laws  define a procedures for determining and mitigating harm after a technology is distributed to consumers. Medical malpractice laws that determine fault when patients are harmed are examples of \textit{ex post} regulations. Further, \textit{ex ante} and \textit{ex post} regulations can interact with each other. Under previous FDA regulations, achieving Class III FDA approval of a medical technology could shield a company from certain degrees of liability claims if a consumer was harmed \cite{FDA_Liability}. In this way, \textit{ex ante} and \textit{ex post} laws change the AI value chain from design and development to deployment. When creating a new AI-based technology, developers and companies must consider both \textit{ex ante} and \textit{ex post} laws in their designs and verification processes.

Lawyers’ roles as AI Creators are reflected in their robust conversations around how to create laws that direct how AI systems are designed, deployed, and developed to ensure the safe and effective use of AI systems. This includes discussions around the kinds of datasets should be allowed for training models \cite{GerkeEtAl2020}, protocols and system requirements when evaluating systems before and during deployment \cite{GerkeEtAl2020}, how governance and maintenance of these systems should be distributed \cite{Price2022}, required methods for validation in localized settings \cite{AbramoffEtAl2020}, and the responsibility of individual care providers in training and using AI systems \cite{Chan2021, MalihaEtAl2021}. 

Limited work has investigated how to coordinate work between legal and technical experts to create effective policies and systems that together support contestability. Within the HCI community, some conversations draw attention to lawyers’ roles as AI Creators in creating policy and laws to enforce ethical design principles and ensure contestability. Alfrink et al. \cite{AlfrinkEtAl2023_Framework} emphasize the need for \textit{ex ante} safeguards to ensure contestability, placing policy level constraints to protect against potential harms through requirements on certifications, including specific aspects of design or requiring certain outputs that enable monitoring and evaluation. Others use the term ``policy-software-decisions loop'' to describe the direct impact of policy on the design of software and subsequently how decisions are made with and around AI systems \cite{AlfrinkEtAl2023_ContestableCars}. Unfortunately, lawyers have been shown to lack understanding around AI systems' capabilities \cite{GorskiAndRamakrishna2021}, which can impair their ability to create laws to effectively support contestability. Collaboration across the policy-software-decisions loop could help address this, but limited work has explored methods for including and leveraging lawyers' roles as AI Creators to support contestability through the intentional co-design of laws and technology.

\section{Lawyers as Intermediaries}
After AI systems are deployed, lawyers continue to play an important part in shaping their contestability.
Lawyers are integral to interpreting the law and determining legal consequences across the network of people tied to harms from decisions with the AI system. Lawyers examine evidence from a case, integrating it with legal information to construct a narrative that can persuade a judge or jury to act in favor of their client's perspective. For example, in a medical malpractice suit against a physician, a lawyer advocating for the patient might work to build a case clearly depicting the physician's negligence in care. A lawyer for the physician might aim to form and solidify an image of the physician as a professional with domain knowledge, skills, and intentional care without oversight. In both cases, the lawyer serves as a conduit for their client's position, acting as a kind of legal prism influencing subsequent legal consequences. Thus, lawyers not only stand between the legal system and clients---directly impacting legal consequences for clients---but also between clients and other people or institutions who have conflicting interests. Consequently, we use the term \textit{Intermediaries} to describe lawyers' roles in navigating legal structures between institutions and stakeholders, interpreting the connections between the harms from using the AI system and the related network of people tied to its use. 

Within the Contestability Community, attention around lawyers' roles as Intermediaries is growing. Warren and Salehi \cite{WarrenAndSalehi2022} describe the role of public defenders in interpreting AI systems' data and their significant impact on those they represent, ``public defenders are bulwark[s] between a poor person charged with a crime and the most consequential harms of state surveillance.'' Other studies have highlighted how lawyers come alongside those they represent, emphasizing their role in the ``political process'' (i.e. power-related process) of contesting an AI decision \cite{KarusalaEtAl2024}.

Although others may accompany stakeholders in pushing back against AI systems, it is important to specifically design for lawyers' capacities as Intermediaries because of their unique legal position.
The term ``intermediaries'' has been used more widely to describe all who come alongside others to push back against unjust or incorrect decisions, including but not restricted to lawyers, NGO's, and other institutions \cite{KarusalaEtAl2024}. However, lawyers are uniquely positioned with respect to the law and their ability to navigate legal structures to push back against AI systems. Lawyers are trained to cultivate skills that allow them to make effective arguments in court and to navigate legal procedures. One study on Contestability underscored how lawyers' abilities to practically enact contestability through the legal system stems from their ``institutional knowledge of in-house experts, attorney-analyst liaisons, and learning from other public defenders'' \cite{JinAndSalehi2024}. Specific design requirements addressing lawyers' needs as Intermediaries can help ensure AI systems are contestable, but limited work has investigated how to understand or design for how they exercise their unique capacities to advocate for clients.

\section{Designing with Lawyers}
Specific recommendations around opportunities for engaging lawyers in co-designing AI systems and contestability mechanisms  need to be created \cite{JinAndSalehi2024}. There's a growing movement to integrate design thinking in the law to support creative human-centered problem solving around such challenges. Proponents argue that design thinking can help lawyers think ``not what the law is, but what it can be'' \cite{Xu2023} and even advocate for the integration of design thinking into law schools' pedagogy \cite{Xu2023, Ursel2017}. In line with this movement is the field of \textit{Legal Design}, the application of design thinking to assess and recreate legal systems, documents, and services \cite{Haapio2014, Hagan2020}. 

Legal design draws on a variety of human-centered design methods and has been used to support community- and user-centered revisions of legal services \cite{PasseraAndHaapio2013, Haapio2014}. For example, the Wikimedia foundation used Legal Design as a part of re-creating their trademark policy, so they could support cross-field discussion among its lawyers, contributors, and users around the new policy \cite{Haapio2014}. Others have used Legal Design to approach redesigning contracts, so users more easily understand what they are agreeing to \cite{PasseraAndHaapio2013}.

Work in Legal Design demonstrates the potential for using design thinking with lawyers to tackle challenges around technology use. But it has largely been used to innovate around legal changes with new technology, rather than technical methods, and there is a need for co-designing with lawyers to create specific recommendations for the design of AI systems \cite{JinAndSalehi2024}. Because lawyers act as AI Creators and Intermediaries, tailoring design methods for these two roles can open ways to coordinate policy and technical efforts for contestability. Here we discuss several areas where design innovations could support cross-disciplinary work to improve AI systems' designs. We also articulate corresponding challenges to help focus the development and refinement of these design innovations \cite{HirshEtAl2017}.

\subsection{Opportunity to Create Co-Design Methods}
Creating co-design methods that leverage lawyers' roles as AI Creators and Intermediaries can coordinate work between the communities, expanding the design space by aligning technical and legal pathways  for contestability. Research has recognized the need not just to investigate technical methods for contestable AI systems, but the importance of laws in enforcing them to ensure ethical and safe designs \cite{AlfrinkEtAl2023_Framework}. In order to coordinate work between the lawyers and other AI Creators, there is a need for cross-disciplinary communication \cite{OtterloAndAtzmueller2018, AlfrinkEtAl2023_Framework} to ``continually assessing the changing risk environment'' and address it \cite{Oswald2013}. Co-design methods can provide space for experts in diverse fields to contribute and share knowledge \cite{VinesEtAl2013}, supporting this kind of cross-disciplinary communication to improve contestability.

Co-design methods that recognize lawyers' roles as AI Creators and Intermediaries can enhance the design of both laws and technology to enable actionability on several fronts. First, they can help developers understand how laws translate to technical requirements for AI systems, so they can better anticipate and design for their impact on the sociotechnical environment. For example, co-design methods could help with tailoring AI explanations for requirements around contestability \cite{LeofanteEtAl2024}, ensuring AI systems have the legally required explanations.

Co-design methods can also help ensure laws have their intended effect on technology design. For example, some have highlighted the role that speculative co-design practices could play in understanding the potential impact of technologies and discussing the policy dimensions of their implementation \cite{SpaaEtAl2019, WarrenAndSalehi2022}. While speculative design holds a lot of promise, it is important to account for ``issue spotting''---a skill taught in law school where lawyers identify specific features of a case that are legally problematic \cite{CuthillAndMcCartney1993}. Without carefully considering what kinds of details are included in speculative activities, lawyers will likely turn to issue spotting, leading them to fixate on details in the interfaces, rather than thinking about the broader context in which the system is used and how that connects to users' and current challenges in the law. But well-tailored descriptions, visuals, and prototypes can help lawyers generate ideas without turning to issue spotting and design more effective laws.

Co-design methods can also help pinpoint new areas for innovation beyond explanations. Lawyers act as Intermediaries representing those impacted by decisions made with AI systems. Prior work has shown how challenges with technology tied to AI systems can undermine lawyers' legal defense, translating to significant consequences for clients and negatively impacting the contestability of AI systems \cite{WarrenAndSalehi2022}. Consequently, there's a need to address barriers in engaging lawyers in ``evaluation design and building tools'' \cite{JinAndSalehi2024}. Co-design methods with lawyers could meet these calls and clarify AI systems' designs to more clearly communicate rights and liabilities when using a system. This could help users and lawyers more easily understand and pursue avenues for change when adverse outcomes happen.

\subsection{Creating Boundary Objects}
The other design opportunity is to create design artifacts (e.g. visualizations, graphics, etc) that can serve as \textit{boundary objects}, promoting shared understandings for legal and technical communities. Boundary objects help communicate shared articulations and common understandings between knowledge communities that have different cultures. 
Boundary objects have been used in both the HCI and Legal Design Communities to help critique and innovate around infrastructures. In HCI, boundary objects are central to coorperative and participatory forms of design \cite{VinesEtAl2013} and have been used to critique technology \cite{VinestEtAl2012} and direct new avenues for research \cite{SpaaEtAl2019}. 

Similarly, in Legal Design, boundary objects have been used in community-centered work between researchers institutions, lawyers, and clients to re-make policies in response to new uses of technology. For example, the Wikimedia Foundation collaborated with researchers to create a set of boundary objects to communicate about and elicit feedback from the Wikimedia community about the current and proposed trademark policy. These boundary objects took the form of graphics that used plain language and visualization techniques to clearly communicate how trademarks might differ between the policies. It pinpointed the most salient aspects of different trademark uses, and then linked back to the legal text for additional information. In this way, it helped people understand the different actions they could take with the trademarks and how they would differ between the policies, so they could provide feedback on changes for the new policy. 

We advocate for the continued and expanded use of boundary objects to communicate with lawyers throughout the co-design process and to summarize findings from collaborations between the communities. Doing so can help expand and develop best practices around designing AI systems and promote innovation around contestability. For example, a boundary object could take the form of a study artifact that is annotated to support others in creating their own inter-disciplinary co-design methods; prototypes of systems could be used to generate discussions around legal or technical challenges; and visual summaries of findings could help others more easily tackle outstanding challenges around state-of-the-art contestability findings.

\subsection{Differences that Co-Design Methods and Boundary Objects Must Navigate}

There are several differences between technical and legal communities that co-design methods and boundary objects must navigate. Here we articulate two challenges to help focus the development and refinement of these methods: (1) lawyers potential lack of understanding around AI systems' capabilities; and (2) fundamental terminological differences between the communities that may cause confusions. 

First, lawyers may not understand how the technological capabilities of AI systems relate to contestability. For example, many legal scholars argue that more transparent, inherently explainable systems translate to trust and accountability \cite{BaitAndSwitch2023, Doshi-VelezEtAl2017, Ehsan2024}, but they have incorrect expectations around what explainability methods can do or how they contribute to transparency. One set of researchers reported that lawyers, when working with an explainable AI tool, expected the system to support its conclusions with citations and facts relevant to the system's decision \cite{GorskiAndRamakrishna2021}. 
This can be a problematic interpretation of explainability for many high performing AI systems that are not inherently explainable and could not provide this information. 

There are also fundamental terminology differences between the communities. The computer science \cite{MireiaEtAl2023, LeofanteEtAl2024} and legal communities \cite{BaitAndSwitch2023} both see explanations as key mechanisms for contestability. However, they do not agree on the definition of ``explanations''. 
In the computer science community, there is no overarching definition of ``explainability'' \cite{NyrupAndRobinson2022}. Explainable, interpretable, intelligible, and scrutable \cite{AbdulEtAl2018, KhosraviEtAl2022} have been used across the field to describe a variety of overlapping approaches to AI systems. Even among computer scientists, the two most widely used terms---``explainable'' and ``interpretable''---are criticized for the ambiguity with which they are used, described, and evaluated \cite{NyrupAndRobinson2022, Lipton2018}. 

On the other hand, legal scholars are defining their own ways of using and distinguishing explainability approaches. For instance, Babic and Cohen \cite{BaitAndSwitch2023} define ``interpretable'' algorithms in a way that maps to inherently explainable AI algorithmic approaches, and they use ``explainable'' to refer to post-hoc methods of explainability. %
In direct contrast, Abdul et al. \cite{AbdulEtAl2018}, a group of computer science researchers, use ``interpretable machine learning (iML)'' to refer to post-hoc methods of explainability. Later the same paper uses ``Explainable AI'' as a term encompassing iML---a double conflict of terminology with Babic and Cohen. When developing co-design methods and creating boundary objects, it is important to take into account these fundamental differences in terminological standards. Differences can significantly impact the participation of legal participants and can cause confusion as to what regulations and proposed policies mean for the development of AI applications. 

\section{Conclusions}
Laws and regulations dramatically shape the creation of AI systems. However, lawyers' roles in the AI value chain remains unrecognized. Recognizing lawyers as AI Creators and Intermediaries pinpoints areas of co-design that can improve and expand the contestability of AI systems. 
Our work contributes a first step in illuminating the specific ways in which lawyers' impact the AI value chain as AI Creators and Intermediaries, and in advocating for cross-disciplinary design with lawyers. We articulate challenges in engaging cross-disciplinary design with lawyers and suggest new opportunities for design innovations to meet this challenges, leverage lawyers' skills, and increase the contestability of AI systems.

\begin{acks}
This material is based upon work supported by the National Science Foundation GRFP under Grant No. DGE-2039655. Any opinion, findings, and conclusions or recommendations expressed in this material are those of the authors(s) and do not necessarily reflect the views of the National Science Foundation.
\end{acks}

\bibliographystyle{ACM-Reference-Format}
\bibliography{main}

%\appendix

%\section{If Needed}

\end{document}